\documentclass[journal]{IEEEtran}
\usepackage{amsmath,amsfonts}
\usepackage{algorithmic}
\usepackage{algorithm}
\usepackage{array}
\usepackage{textcomp}
\usepackage{stfloats}
\usepackage{url}
\usepackage{verbatim}
\usepackage{graphicx}
\usepackage{subfigure}
\usepackage{cite}
\usepackage{multicol}
\usepackage{color}

\begin{document}
\title{Global in Local: A Convolutional Transformer for SAR ATR FSL}
%
%
%

\author{Chenwei Wang,~\IEEEmembership{Student Member,~IEEE,}
        Yulin Huang,~\IEEEmembership{Senior Member,~IEEE,}
        Xiaoyu Liu,~\IEEEmembership{Student Member,~IEEE,}
        Jifang Pei,~\IEEEmembership{Member,~IEEE,}
        Yin Zhang,~\IEEEmembership{Member,~IEEE,}
        Jianyu Yang,~\IEEEmembership{Member,~IEEE}
\thanks{This work was supported by the National Natural Science Foundation of China under Grants 61901091 and 61901090. (\emph{Corresponding author: Yulin Huang.})

       The authors are with the Department of Electrical Engineering, University of Electronic Science and Technology of China, Chengdu 611731, China (e-mail: yulinhuang@uestc.edu.cn; dbw181101@163.com).}}

\maketitle

\begin{abstract}
Convolutional neural networks (CNNs) have dominated the synthetic aperture radar (SAR) automatic target recognition (ATR) for years. However, under the limited SAR images, the width and depth of the CNN-based models are limited, and the widening of the received field for global features in images is hindered, which finally leads to the low performance of recognition. To address these challenges, we propose a Convolutional Transformer (ConvT) for SAR ATR few-shot learning (FSL). The proposed method focuses on constructing a hierarchical feature representation and capturing global dependencies of local features in each layer, named global in local. A novel hybrid loss is proposed to interpret the few SAR images in the forms of recognition labels and contrastive image pairs, construct abundant anchor-positive and anchor-negative image pairs in one batch and provide sufficient loss for the optimization of the ConvT to overcome the few sample effect. An auto augmentation is proposed to enhance and enrich the diversity and amount of the few training samples to explore the hidden feature in a few SAR images and avoid the over-fitting in SAR ATR FSL. Experiments conducted on the Moving and Stationary Target Acquisition and Recognition dataset (MSTAR) have shown the effectiveness of our proposed ConvT for SAR ATR FSL. Different from existing SAR ATR FSL methods employing additional training datasets, our method achieved pioneering performance without other SAR target images in training.
\end{abstract}

\begin{IEEEkeywords}
SAR ATR, FSL, convolutional transformer, hybrid loss, auto data augmentation
\end{IEEEkeywords}

%
\IEEEpeerreviewmaketitle

\section{Introduction}

\IEEEPARstart{S}{AR} is an important microwave remote sensing system in both mechanism and application\cite{intro1,wang2022semi,wang2020deep,wang2022sar,wang2021deep}, which makes SAR automatic target recognition (ATR) become one of the most important and crucial issues in SAR practical application. In recent several years, deep learning has also illustrated its effectiveness in the field of SAR ATR. Many deep learning-based methods are proposed in SAR ATR applications and achieved remarkable results \cite{intro3,intro5,wang2023entropy,wang2023sar,wang2021multiview,wang2019parking,wang2020multi}.

However, to acquire great generalization performance of recognition under various imaging scenarios, these existing SAR ATR algorithms require abundant labeled samples for each target type to train the deep network. However, it is often impossible to provide sufficient images in practical applications. These problems have promoted the researches on few-shot learning (FSL) in SAR ATR, which can be divided into two categories: data augmentation methods and deep model-based methods \cite{add1,add2, add3,EliMRec,SLMRec,BundleGT,DBLP:conf/emnlp/ZhaoWLSZ022,DBLP:conf/acl/Zhao00WZZC23,wang2022recognition,wang2022global}.

Data augmentation method is a technique to enhance the amount and enrich the quality of training SAR images with the help of sufficient similar SAR images. For example, Wang et al. \cite{intro2} designed a semi-supervised learning framework including self-consistent augmentation rule, mixup-based labeled and unlabeled mixture, and weighted loss, to utilize unlabeled data during training. Deep model-based methods mainly depend upon architecture design for rapid generalization of few-shot learning tasks. For example, Wang et al. \cite{intro5} proposed a hybrid inference network (HIN) including an embedding network stage and a hybrid inference strategy stage, which obtained good performance of 3-target classification.

\begin{figure*}[htb]
\centering
\centering\includegraphics[width=0.9\textwidth]{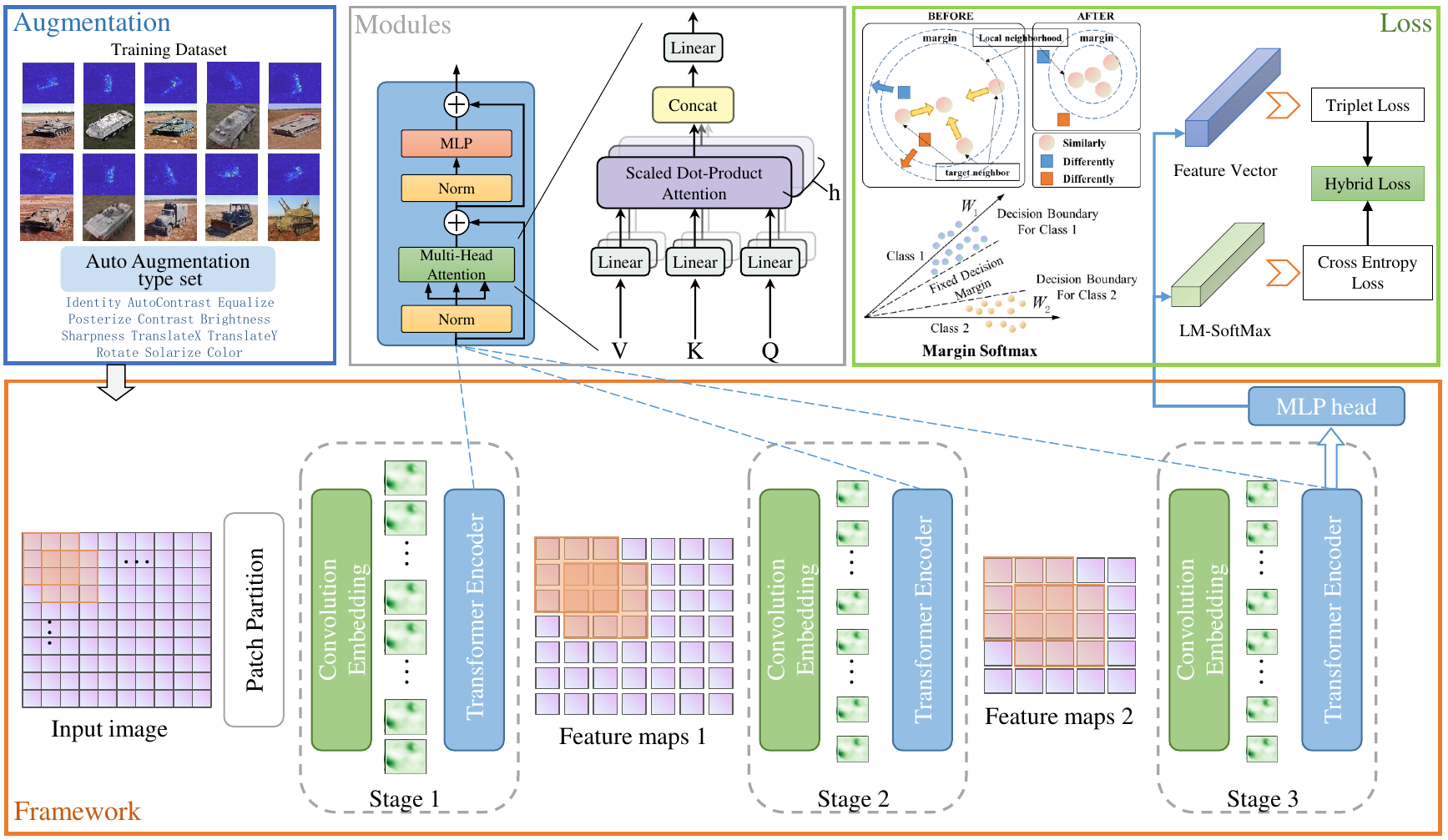}
\caption{Framework of the proposed ConvT.}
\label{Framework}
\end{figure*}

These data augmentation methods construct the manifold structure of unlabeled samples similar to few labeled training samples. However, these methods still acquire the similar manifold structure of sufficient unlabeled samples. The sufficient unlabeled samples do not always exist, let alone in the scene of changing characteristics in SAR images under various imaging scenarios. These deep model-based methods exploit sufficient labeled training samples under quite similar imaging conditions to train the models. However, these model-based methods still need sufficient SAR images to avoid overfitting and acquire great generalization of recognition performance.

The key to addressing the problem of insufficient SAR samples in SAR ATR needs two critical elements: an effective framework to extract optimal features, and a hybrid optimization for limited SAR images.
The framework can extract effective feature for the recognition with a relatively shadow structure. The hybrid optimization can exploit the sufficient information for the optimization of the framework to overcome the limited sample effect.
In light of the vigorous development and superior performance of visual transformers, it can provide a novel view for SAR image interpretation different from the convolutional neural networks (CNNs). As opposed to CNNs that mainly focus on extracting local features, a transformer can capture global dependencies in the image, which is mainly based on the self-attention mechanism with strong representation capabilities \cite{intro11}. For SAR ATR FSL, limited SAR images hinder the width and depth of the CNN-based model, which also hinder widening the received field for great generalization of recognition performance and achieving high performance of recognition. These limitations of CNN-based backbone may be the reason that these existing SAR ATR methods for FSL still need sufficient training data to support the CNN model \cite{intro11}.

Therefore, we proposed a Convolutional Transformer (ConvT) for SAR ATR FSL, which naturally integrates the local features of CNNs and global dependencies of transformers with a novel hybrid loss for the limited SAR images. Our model consists of three parts as shown in Fig.1, convolutional transformer (orange box), hybrid loss (green box), and auto data augmentation (blue box). The pipeline of our model can be described as follow. Limited SAR images go through auto data augmentation to enhance its amount and diversity. Local and global features are extracted and interpreted by convolutional transformer, and the hybrid loss of label propagation and contrastive learning is proposed to provide sufficient optimization for better generalization of recognition performance. The main contributions are as follows.

i) We proposed a convolutional transformer, which constructs a hierarchical feature representation and extracts the global dependency of local features in each layer. This model not only preserves the capability of CNNs to extract local features and widen the received field in a hierarchical structure but also acquires the global dependencies in the whole SAR images and local features which provides better generalization for recognition under limited training samples.

ii) The hybrid loss is proposed for providing more abundant optimization for the proposed model. The hybrid loss can overcome the few sample effect by interpreting the few SAR images in the forms of recognition labels and contrastive image pairs. The auto data augmentation is proposed to express many augmentation policies for each epoch in training to explore the hidden feature in a few SAR images.

iii) The proposed model achieves competitive performance to state-of-the-art methods on standard benchmarks. Without any other SAR target images in training except k-shot for each class (support samples), the recognition rates of 10 support samples for each class are above 85.00\%, and the rates of 5 support samples for each class are above 75.00\%.

The remaining structure of this paper is organized as follows. The proposed method is described in Section \uppercase\expandafter{\romannumeral2}. Section \uppercase\expandafter{\romannumeral3} illustrates the experiments and results. Finally, a brief conclusion is drawn in Section \uppercase\expandafter{\romannumeral4}.

\section{Proposed Method}

To address the FSL problem in SAR ATR, and overcome the limitation of the CNN-based backbone for SAR ATR, the proposed method consists of three main components: 1) a convolutional transformer. 2) a hybrid loss. 3) an auto augmentation. The three parts are introduced in detail as follows.

\subsection{Framework of ConvT}

The proposed ConvT aims to overcome the limitation of the CNN-based backbone for SAR ATR FSL. It naturally integrates convolutional layers and transformers to construct a hierarchical feature representation that can extract local features layer by layer and focuses on global dependencies of local features in each layer and the whole image. 

The whole framework of ConvT is constructed in several stages by stacking convolution embedding and transformer encoder as shown in Fig.1. The input image is split into non-overlapping patches by a patch partition module. Each patch is one part of the whole image and all the patches can restructure back into the original images. In each stage, there are two parts. First, these patches are reshaped to the 2D spatial grid and go through convolution embedding to extract local features. This convolution embedding can allow each stage to widen the received field and reduce the feature resolution progressively to acquire spatial downsampling and boost the density of features. Second, a transformer encoder is employed to capture global dependencies of local features in each stage with the position embedding. Finally, through several stages as above, the feature maps of the last transformer encoder are input into one MLP layer to predict the class. The iterative process in ConvT can be formulated as
\begin{equation}
    {I_i} = conv\left( {{T_{i - 1}}} \right)
\end{equation}
\begin{equation}
    {L_i} = MHA\left( {LN\left( {{I_i}} \right)} \right){\rm{ + }}{I_i}
\end{equation}
\begin{equation}
    {T_i} = MLP\left( {LN\left( {{L_i}} \right)} \right) + {L_i}
\end{equation}
where ${{T_{i - 1}}}$ and ${{T_{i}}}$ mean the output feature maps of $i-1$ stage and $i$ stage, $conv$ means the convolution embedding, ${LN}$ and $MHA$ denote the layernorm and multi-head self attention, $MLP$ means multilayer perceptron layer.

The structure of the transformer encoder is shown in Fig.1. After the layer normalization, the feature maps are transformed into three matrixes, $Q$, $K$ and $V$. Then the three matrixes go through the multi-head attention which is shown in Modules of Fig.1. For a more comprehensive exploration of the target information, the attention employed different linear transformations to the features and calculate the scaled dot-product attention as
\begin{equation}
    attention = {\mathop{\rm softmax}\nolimits} \left( {\frac{{Q{K^T}}}{{\sqrt {{d_k}} }}} \right)V
\end{equation}
where ${d_k}$ is the scaled parameter. The outputs of the scale dot-product attention are concatenated and linearly transformed. Another layer normalization and one  MLP layer are employed after the addition. Besides, the layer normalization is a horizontal normalization which comprehensively considers the inputs of all dimensions in one layer. It calculates the mean and variance of the inputs in one layer and employs the same normalization operation in the inputs of all dimensions.

Through the designed framework above, the model not only retains the hierarchical feature representation of CNNs, but also captures the global dependencies in local features of each stage and whole image, which has fewer demands for deep structures and acquires high potential for better recognition performance than CNNs.

The computational complexity of the proposed method is in the order of $O\left( {{{\hat{V}}}_{1}}\cdot {{{\hat{V}}}_{2}}\cdot {{W}_{1}}\cdot {{W}_{2}}\cdot F\cdot I \right)$, where ${{\hat{V}}_{1}}\cdot {{\hat{V}}_{2}}$ is the size of input images, ${{W}_{1}}\cdot {{W}_{2}}$ is the size of convolution kernel, $F$ and $I$ are the numbers of the convolution kernels and feature maps. Therefore, the complexity of our method is in the same order as that of other deep learning methods.

To provide enough optimization gradient and overcome the few sample effect, hybrid loss and auto augmentation are proposed as follows.

\begin{figure}[htb]
\centering
\centering\includegraphics[width=0.48\textwidth]{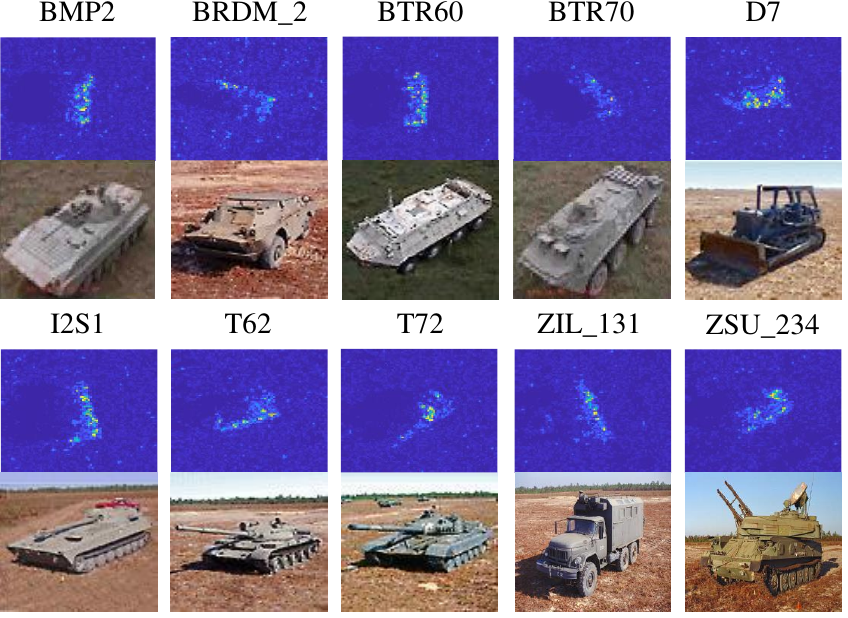}
\caption{Optical images and corresponding SAR images of ten classes of objects in the MSTAR dataset.}
\label{dataset}
\end{figure}

\begin{table}[!htb]
\renewcommand{\arraystretch}{1.2}
\setlength\tabcolsep{7pt}
\centering
\caption{Original Images in MSTAR Dataset Under SOC}
\label{SOCdataset}
\begin{tabular}{lccccc}
\hline\hline
Target Type & BMP2 & BRDM2 & BTR60 & BTR70 & D7 \\ \hline
Training($17^{\circ}$) & 233 & 298 & 256 & 233 & 299 \\
Testing($15^{\circ}$) & 195 & 274 & 195 & 196 & 274 \\ \hline\hline
Target Type & 2S1 & T62 & T72 & ZIL131 & ZSU235 \\ \hline
Training($17^{\circ}$) & 299 & 299 & 232 & 299 & 299 \\
Testing($15^{\circ}$) & 274 & 273 & 196 & 274 & 274 \\ \hline\hline
\end{tabular}
\end{table}

\subsection{Hybrid loss and Auto Augmentation}

To tackle the insufficient optimization of the FSL problem in SAR ATR, we proposed hybrid loss consisting of cross entropy loss and triplet loss to construct abundant positive and negative image pairs in one batch and provide sufficient loss for the optimization of the ConvT. 
When the SAR images go through the framework, and the MLP layer gives the predicted probability vectors of recognition results, the hybrid loss is calculated.
The visual interpretation is shown in Fig.1, the cross entropy loss is calculated after LM-SoftMax \cite{method1} to expand the inter-class gap, and the triplet loss \cite{method2} has a margin to expand the desired difference between the anchor-positive distance and the anchor-negative distance. In our method, the LM-SoftMax-based cross entropy loss and triplet loss are calculated as

\begin{equation}
{L_e}\left( {{\bf{w,b}}} \right) = - \sum\limits_{i = 1}^C {{y_i}\log \left( {p\left( {{y_i}\left| {{{\bf{x}}}} \right.} \right)} \right)}
\end{equation}
\begin{equation}
{L_t} = \max \left( {d\left( {a,p} \right) - d\left( {a,n} \right) + {\rm{margin}},0} \right)
\end{equation}
\begin{equation}
    {L_b} = {L_e} + {L_t}
\end{equation}
where ${{p}\left( {{{y_i}}\left| {{{\bf{x}}}} \right.} \right)}$ is the probability vector of the recognition result of the $i$th SAR chip, ${y_i}$ is the recognition labels and $C$ is the number of the recognition classes, $d$ and $\rm{margin}$ mean the distance and the desired difference between the anchor-positive distance and the anchor-negative distance.

\begin{table}[htb]
\renewcommand{\arraystretch}{1.2}
\setlength\tabcolsep{15pt}
\caption{Training and Testing dataset under EOCs}
\centering
\label{EOCdataset}
\begin{tabular}{lccc}
\hline\hline
Train      & Number & Test(EOC-D) & Number \\ \hline
2S1        & 299    & 2S1(b01)    & 288    \\ 
BRDM2      & 298    & BRDM2(E71)  & 287    \\ 
T72        & 232    & T72(A64)    & 288    \\ 
ZSU234     & 299    & ZSU234(d08) & 288    \\ \hline\hline
Train      & Number & Test(EOC-CV) & Number \\ \hline
BMP2       & 233    & T72(S7)     & 419    \\ 
BRDM2      & 298    & T72(A32)    & 572    \\ 
BTR70      & 233    & T72(A62)    & 573    \\ 
T72        & 232    & T72(A63)    & 573    \\ 
           &        & T72(A64)    & 573    \\ \hline\hline
Train      & Number & Test(EOC-VV) & Number \\ \hline
BMP2       & 233    & T72(SN812)  & 426    \\ 
           &        & T72(A04)    & 573    \\ 
BRDM2      & 298    & T72(A05)    & 573    \\ 
           &        & T72(A07)    & 573    \\ 
BTR70      & 233    & T72(A10)    & 567    \\ 
           &        & BMP2(9566)  & 428    \\ 
T72        & 232    & BMP2(C21)   & 429    \\ \hline\hline
\end{tabular}
\end{table}

Auto augmentation aims to enhance and enrich the diversity and amount of the few training samples to explore the hidden feature in a few SAR images and avoid the over-fitting in SAR ATR FSL. It consists of two main stages with one transformation set. Given $N$ available transformations in the transformation set, the transformation number $K$, global distortion $D$, and threshold for auto augmentation and each transformation in one epoch, ${M_a}$ and ${M_{each}}$. First, for each epoch in training, a number $m$ conforming to the standard normal distribution is generated. if $m \ge {M_a}$, the augmentation is utilized in this epoch. Then $K$ transformations are randomly chosen from the $N$ available transformations in the transformation set. $K$ numbers conforming to standard normal distribution are generated $\left\{ {{m^1},{m^2},{m^3} \ldots ,{m^k}} \right\}$, if ${m^i} \ge {M_{each}}$, then ${i}{\rm{th}}$ transformer of the randomly chosen $K$ transformations is applied in this epoch. Therefore, auto augmentation may thus express ${N^K}$ potential policies for each epoch in training to explore the hidden feature in a few SAR images and increase regularization strength.

\section{Experimental Results}

\begin{table*}[htb]
\renewcommand{\arraystretch}{1.2}
\setlength\tabcolsep{10pt}
\caption{Results of FSL Algorithms under SOC and EOCs}
\centering
\label{recognitionperformance}
\begin{tabular}{lccccc}
\hline\hline
\multicolumn{6}{c}{SOC}                                                                                                                 \\ \hline
Algorithms           & 10-way 1-shot        & 10-way 2-shot        & 10-way 5-shot        & 10-way 10-shot       & 10-way 25-shot       \\ \hline
DeepEMD \cite{camp1}              & 36.19$\pm$0.46       & 43.49$\pm$0.44       & 53.14$\pm$0.40       & 59.64$\pm$0.39       & 59.71$\pm$0.31       \\ 
DN4 \cite{camp2}                 & 33.25$\pm$0.49       & 44.15$\pm$0.45       & 36.19$\pm$0.46       & 36.19$\pm$0.46       & 36.19$\pm$0.46       \\ 
Prototypical Network \cite{camp3} & 40.94$\pm$0.47       & 54.54$\pm$0.44       & 36.19$\pm$0.46       & 36.19$\pm$0.46       & 36.19$\pm$0.46       \\ 
DKTS-N  \cite{intro6}             & \textbf{49.26$\pm$0.48}       & \textbf{58.51$\pm$0.42}       & 72.32$\pm$0.32       & 84.59$\pm$0.24       & 96.15$\pm$0.08       \\ 
Ours                 & 42.57$\pm$0.79       & 54.37$\pm$0.62       & \textbf{75.16$\pm$0.21 }      &\textbf{ 88.63$\pm$0.22}      & \textbf{96.52$\pm$0.15 }      \\ \hline
\multicolumn{6}{c}{EOC-D}                                                                                                               \\ \hline
Algorithms           & 4-way 1-shot         & 4-way 2-shot         & 4-way 5-shot         & 4-way 10-shot        & 4-way 25-shot       \\ \hline
DeepEMD \cite{camp1}             & 56.81$\pm$0.99       & 62.80$\pm$0.78       & 36.19$\pm$0.46       & 36.19$\pm$0.46       & 36.19$\pm$0.46       \\ 
DN4 \cite{camp2}                   & 46.59$\pm$0.83       & 51.41$\pm$0.69       & 36.19$\pm$0.46       & 36.19$\pm$0.46       & 36.19$\pm$0.46       \\ 
Prototypical Network \cite{camp3} & 53.59$\pm$0.93       & 56.57$\pm$0.53       & 36.19$\pm$0.46       & 36.19$\pm$0.46       & 36.19$\pm$0.46       \\ 
DKTS-N \cite{intro6}               &  \textbf{61.91$\pm$0.91}       & 63.94$\pm$0.73       & 67.43$\pm$0.48       & 71.09$\pm$0.41    & 78.94$\pm$0.31       \\ 
Ours                 &59.57$\pm$0.76 & \textbf{64.06$\pm$0.88} &\textbf{68.17$\pm$0.38}      &\textbf{74.80$\pm$0.20}  & \textbf{79.14$\pm$0.42}                   \\ \hline
\multicolumn{6}{c}{EOC-CV}                                                                                                              \\ \hline
Algorithms          & 4-way 1-shot         & 4-way 2-shot         & 4-way 5-shot         & 4-way 10-shot        & 4-way 25-shot       \\ \hline
DeepEMD \cite{camp1}              & 38.39$\pm$0.86       & 45.65$\pm$0.75       & 36.19$\pm$0.46       & 36.19$\pm$0.46       & 36.19$\pm$0.46       \\ 
DN4 \cite{camp2}                  & 46.13$\pm$0.69       & 51.21$\pm$0.62       & 36.19$\pm$0.46       & 36.19$\pm$0.46       & 36.19$\pm$0.46       \\ 
Prototypical Network \cite{camp3} & 43.59$\pm$0.84       & 51.17$\pm$0.78       & 36.19$\pm$0.46       & 36.19$\pm$0.46       & 36.19$\pm$0.46       \\ 
DKTS-N \cite{intro6}               & \textbf{47.26$\pm$0.79} & \textbf{53.61$\pm$0.70} & 62.23$\pm$0.56 & 68.41$\pm$0.51       & 74.51$\pm$0.36       \\ 
Ours                 & 44.32$\pm$0.65       & 51.93$\pm$0.82       & \textbf{64.12$\pm$0.34} & \textbf{89.74$\pm$0.18} & \textbf{90.95$\pm$0.23}       \\ \hline
\multicolumn{6}{c}{EOC-VV}                                                                                                              \\ \hline
Algorithms           & 4-way 1-shot         & 4-way 2-shot         & 4-way 5-shot         & 4-way 10-shot        & 4-way 25-shot       \\ \hline
DeepEMD \cite{camp1}              & 40.92$\pm$0.76       & 49.12$\pm$0.65       & 36.19$\pm$0.46       & 36.19$\pm$0.46       & 36.19$\pm$0.46       \\ 
DN4 \cite{camp2}                  & 47.00$\pm$0.72       & 52.21$\pm$0.61       & 36.19$\pm$0.46       & 36.19$\pm$0.46       & 36.19$\pm$0.46       \\ 
Prototypical Network \cite{camp3} & 45.13$\pm$0.72       & 52.86$\pm$0.65       & 36.19$\pm$0.46       & 36.19$\pm$0.46       & 36.19$\pm$0.46       \\ 
DKTS-N \cite{intro6}               & \textbf{48.91$\pm$0.70}       & 55.14$\pm$0.58       & 65.63$\pm$0.49       & 70.18$\pm$0.42       & 76.97$\pm$0.35       \\
Ours                 & 42.27$\pm$0.89  &\textbf{58.27$\pm$0.68} & \textbf{68.05$\pm$0.52} &\textbf{83.55$\pm$0.25} &\textbf{91.98$\pm$0.31}\\ \hline\hline
\end{tabular}
\end{table*}

In this section, the extensive experiments are conducted under both the Standard Operating Condition (SOC) and the Extended Operating Condition (EOC) consisting of EOC- CV (configuration variant), EOC-D (depression variant) and EOC-VV (version variant). N-way K-shot denotes the K training samples for all N target classes.

\subsection{Dataset}
The MSTAR dataset is a benchmark dataset for the SAR ATR performance assessment. The dataset contains a series of $0.3m\times0.3m$ SAR images of ten different classes of ground targets. The optical images and corresponding SAR images of ten classes of targets in the MSTAR dataset are shown in Fig.\ref{dataset}. The training and testing data under SOC and EOCs have been shown in Table \ref{SOCdataset} and Table \ref{EOCdataset}. The proposed method is tested and evaluated on a computer with Intel Core I7-9700K at 3.6GHz CPU, Gefore GTX 1080ti GPU with two 16GB memories with the open-source PyTorch framework. In our experiments, the parameters of the auto augmentation are set as below. The available transformations $N$ is 12 as shown in Fig.1, global distortion $D$ is 3, ${M_a}$ and ${M_{each}}$ are 0. For 1-shot and 2-shot, the transformation number $K$ is 5. For other N-shot, the transformation number $K$ is 3.

\subsection{Recognition Performance and Comparison}

This subsection presents the recognition performance and comparison under SOC and EOCs. Other algorithms \cite{camp1,camp2,camp3,intro6} are compared with our ConvT. The recognition ratios of the other algorithms are cited from \cite{intro6}. For K-shot, our method randomly chooses K images for each class in MSTAR to train the model. Other methods also randomly choose K images for each class in MSTAR.

In Table \ref{recognitionperformance}, the recognition performances of SOC are present. Ten-way K-shot involved in the experiments were randomly selected with a $17^{\circ}$ depression angle from the SOC training dataset. The average recognition ratios are calculated after 20 experiments. From Table \ref{recognitionperformance}, it is clear that when the training samples increase, the recognition performance is improved gradually. For the experiments of 25-shot, 10-shot and 5-shot, our ConvT can achieve the highest recognition ratios, 96.52\% for 25-shot, 88.63\% for 10-shot and 75.16\% for 5-shot. Under 10-shot and 5-shot, our ConvT has more obvious advantages than other methods. For the experiments of 1-shot and 2-shot, our method is closer to the best recognition performance of DKTS-N, 54.37\% for 2-shot and 42.57\% for 1-shot. 

In Table \ref{recognitionperformance}, the recognition performances of EOCs are present. Four-way K-shot involved in the experiments were randomly selected from the EOCs’ training dataset. The average recognition ratios are calculated after 20 experiments. From Table \ref{recognitionperformance}, the recognition performance of EOC-D has shown the effectiveness of the capability of extracting global features at the local image level. The recognition ratio of EOC-D under 25-shot, 10-shot, 5-shot, 2-shot and 1-shot are 79.14\%, 74.80\%, 68.17\%, 64.06\% and 59.57\% respectively. The recognition performance decreased lightly when the training samples is reduced. For EOC-CV and EOC-VV, the recognition performances under 25-shot, 10-shot and 5-shot are obviously better than other algorithms, which illustrated our method has a higher upper bound. The recognition ratios of EOC-CV and EOC-VV under 2-shot and 1-shot are close to the best recognition performance from DKTS-N. Besides, the computation time for each image is approximately $15ms$.

From the recognition performance of SOC and EOCs, the results have illustrated that our method has the robustness and effectiveness for SOC and EOCs, the recognition ratios hold at a high level facing the large depression angel, configuration variant and version variant. Without additional training datasets, our method has achieved state-of-the-art performance of SAR ATR FSL.

\section{Conclusion}
For SAR ATR FSL, limited SAR images hinder the width and depth of the CNN-based model, which is the key to wider the receive field of the model for extracting global features in images and achieving high performance of recognition. The proposed ConvT constructed a hierarchical feature representation and extract global features in the local representation of each layer. The hybrid loss constructed abundant anchor-positive and anchor-negative image pairs in one batch and provided sufficient loss for the optimization of the ConvT.
The auto augmentation is employed to enhance and enrich the diversity and amount of the few training samples. Experimental results on the MSTAR dataset have validated the effectiveness and robustness of the proposed ConvT in few-shot recognition in SAR. Different from existing SAR ATR FSL methods employing additional training datasets, our ConvT achieved pioneering performance without other SAR target images in training besides support samples of MSTAR. Though our method achieves high performances in SAR ATR, it still lacks the ability to employ the unlabeld data. In the future work, we will focus on the growing amount of unlabeled data to further improve the robustness of the model and expand the scope of practical applications of SAR ATR.

\bibliographystyle{IEEEtran}
\bibliography{references,ref_add}


%






\end{document}